# Extremal micropolar materials for elastic wave cloaking

Dingxin Sun[1], Yi Chen[2,3]*, Xiaoning Liu[1,4] and Gengkai Hu[1]*

[1]School of Aerospace Engineering, Beijing Institute of Technology, Beijing 100081, China.

[2]Institute of Nanotechnology, Karlsruhe Institute of Technology, Karlsruhe 76128, Germany.

[3]Institute of Applied Physics, Karlsruhe Institute of Technology, Karlsruhe 76128, Germany.

[4]School of Civil Engineering, Changsha University of Science & Technology, Changsha 410114, China.

*Corresponding authors: yi.chen@partner.kit.edu (Y.C.); hugeng@bit.edu.cn (G.H.)

## Abstract

The asymmetric transformation elasticity offers a promising method to control elastic waves. However, this method requires elastic materials that support asymmetric stresses, which is not objective within the linearized Cauchy elasticity for small deformations. Nevertheless, asymmetric stress tensor is a typical feature of micropolar continuum theory. Yet, possible connection between micropolar continuum theory and the asymmetric elasticity transformation has remained elusive. Here, we demonstrate that extremal micropolar media, which refer to micropolar media with easy deformation modes, can be used to design elastic cloaks following the asymmetric transformation method. A metamaterial model is proposed to achieve the required extremal micropolar parameters for cloaking. We further design a two-dimensional metamaterial cloak and verify its cloaking performance numerically. An excellent agreement between the metamaterial cloak simulation and an effective-medium calculation is obtained. This study unveils a novel strategy for controlling elastic waves through micropolar media and also sheds light on interesting properties of extremal micropolar materials.

## Keywords

asymmetric transformation elasticity, extremal micropolar media, easy deformation modes, micropolar metamaterials, elastic cloak

## 1. Introduction

Transformation theory, firstly proposed in electromagnetic wave fields [1], has become a successful design tool for various wave fields, e.g., acoustic waves, optics and so on [2-6]. Many interesting wave devices have been designed based on this theory over the past decade, such as omni-directional absorbers or cloaks [7-9]. However, the application of transformation theory for elastic waves is still challenging due to the complexity of elastic waves, which contain both longitudinal components and transverse components. Transformation elasticity is mainly available for a number of limited situations, e.g., flexure waves in plates or approximation control at high frequency range [10, 11].

Controlling elastic waves via transformation elasticity generally requires two steps. Firstly, the Navier equation for Cauchy materials in a virtual space is transformed to a new governing equation in a physical space [2, 4]. The second step is to realize materials or metamaterials [12-14] that follow the transformed governing equation. For elastic waves, two types of transformed equations exist. In one case, the transformed equation is rather complicate and requires Willis materials [12-14], which requires coupling between stress/momentum and velocity/strain [15]. Therefore, this method is less adopted for designing elastic waves controlling devices. In another version, the transformed equation has the same form as the Cauchy theory, while the fourth order elastic tensor loses its minor symmetry, likewise the stress tensor [4]. The theory is also called asymmetric transformation elasticity [16]. However, the asymmetric elastic theory is not physical or well defined, since the theory doesn't not take into the account of the unbalanced angular momentum caused by asymmetric stress. This problem caused concern on the possibility of designing cloaks following the asymmetric transformation elasticity [17]. At first sight, it seems impossible to design an asymmetric elastic material. Interestingly, the propagation of a small disturbance superimposed onto a largely-deformed hyperelastic material (i.e. the small-on-large theory) experiences an effective asymmetric elastic tensor [18-20]. Therefore, hyperelastic materials, like semi-linear materials, have been theoretically shown to be able to cloak elastic waves [19-21]. Yet, the cloaking effect is limited due to geometry restrictions [19]. The required hyperelastic behavior is also challenging to achieve with artificial microstructures.

Recently, asymmetric elastic materials have been effectively realized by metamaterials [22-28]. In the first type of design, each unit cell contains a mass block that can freely translate but is restrained against rotation by grounded torsional springs [22, 23], or by external magnetic fields [28]. The asymmetric stress is balanced by external torque supplied by the torsional springs or the magnetic fields. To realize such a metamaterial, a complicated mechanism is needed to restrict the rotation but not the translation degrees of freedom (DOFs) of its inner block [24]. In the second type of design, a discrete metamaterial with local rotational resonance is adopted [26]. The asymmetric shear stress expressed by a metamaterial unit cell is compensated by its inner rotation inertia. Both strategies have been adopted to achieve asymmetric elastic tensors required by the asymmetric transformation elasticity. The effectiveness of both metamaterials to mimic asymmetric elastic materials has been confirmed by experiments. We note here that though asymmetric elastic materials and asymmetric elastic theory are not physical [4], yet asymmetric elastic materials can be well approximated by designed metamaterials. This is the key idea of metamaterials [14], achieving effective-medium properties that seem impossible at first sight, e.g., negative index [29, 30].

In this paper, we aim at realizing elastic wave cloaking from a different route, i.e., by using micropolar elastic media, which naturally support asymmetric stress. The advantage is that micropolar elasticity theory is physical and has a rigorous theory foundation, in comparison with the asymmetric elasticity theory. Our initial idea comes from noticing that pentamode materials [31-38] are capable of controlling acoustic waves based on transformation acoustics [39, 40]. Pentamode materials is a typical example of extremal Cauchy materials, which allow easy deformation modes that conceptually cost no strain energy [41-46]. Therefore, we anticipate that special extremal higher-order elastic media can potentially be used to control elastic waves in Cauchy materials [47]. A promising candidate is the micropolar continuum media, also called Cosserat media [48-51], which is relatively simple in mathematics and more importantly can capture asymmetric stresses. We remark that micropolar elasticity has been adopted to study asymmetric stresses in granular mechanics fields or seismology [52-54], even far before the emergence of transformation theory and metamaterials. In our paper, we particularly focus on extremal micropolar materials with a zero higher-order elastic tensor, also termed as reduced Cosserat media [55]. We show that such media indeed provide a new strategy for realizing asymmetric transformation elasticity. We provide guidelines for designing the easy deformation modes required by transformation. We also propose a discrete metamaterial to realize the extremal micropolar continuum parameters. Furthermore, we design a two-dimensional (2D) elastic cloak based on the proposed metamaterial and demonstrate the effectiveness of the theory and the metamaterial by numerical simulations.

The paper is organized as follows. We establish in detail the connection between micropolar continuum theory and asymmetric transformation in Section 2. The required easy deformation modes of extremal micropolar media for transformation are discussed in Section 3. Then, a metamaterial model is proposed and its effective micropolar elastic parameters are derived analytically. The metamaterial is numerically verified by wave simulations. In Section 4, we design a 2D elastic cloak based on the proposed micropolar metamaterial. The cloaking performance is validated numerically and compared with results from effective-medium model. Finally comes the conclusion.

## 2. Micropolar continuum theory for elastic cloaking

### *2.1 Asymmetric transformation elasticity*

We briefly recall the asymmetric transformation elasticity [4] through the design of a 2D annular cloak. A virtual space (Fig. 1(a)) is assumed to be filled with a Cauchy material with the mass density $\rho_0$ and the elastic tensor $C_{0ijkl} = C_{0jikl} = C_{0klij}$. We limit the study in this paper to small deformations and only consider linear elastic waves propagating inside the material. Omitting a time harmonic term $\exp(-\mathrm{i}\omega t)$, with i being the imaginary unit, $\omega$ the angular frequency and $t$ the time, the dynamic equation and constitutive equation for waves propagating in the virtual space in frequency domain write as

$$-\rho_0\omega^2 u_i(\mathbf{X}) = \frac{\partial \sigma_{ji}(\mathbf{X})}{\partial X_j}, \quad \sigma_{ij} = C_{0ijkl}\frac{\partial u_l(\mathbf{X})}{\partial X_k}, \tag{1}$$

in which, $u_i(\mathbf{X})$ and $\sigma_{ij}(\mathbf{X}) = \sigma_{ji}(\mathbf{X})$ represent the displacement and the symmetric stress tensor in the virtual space, respectively.

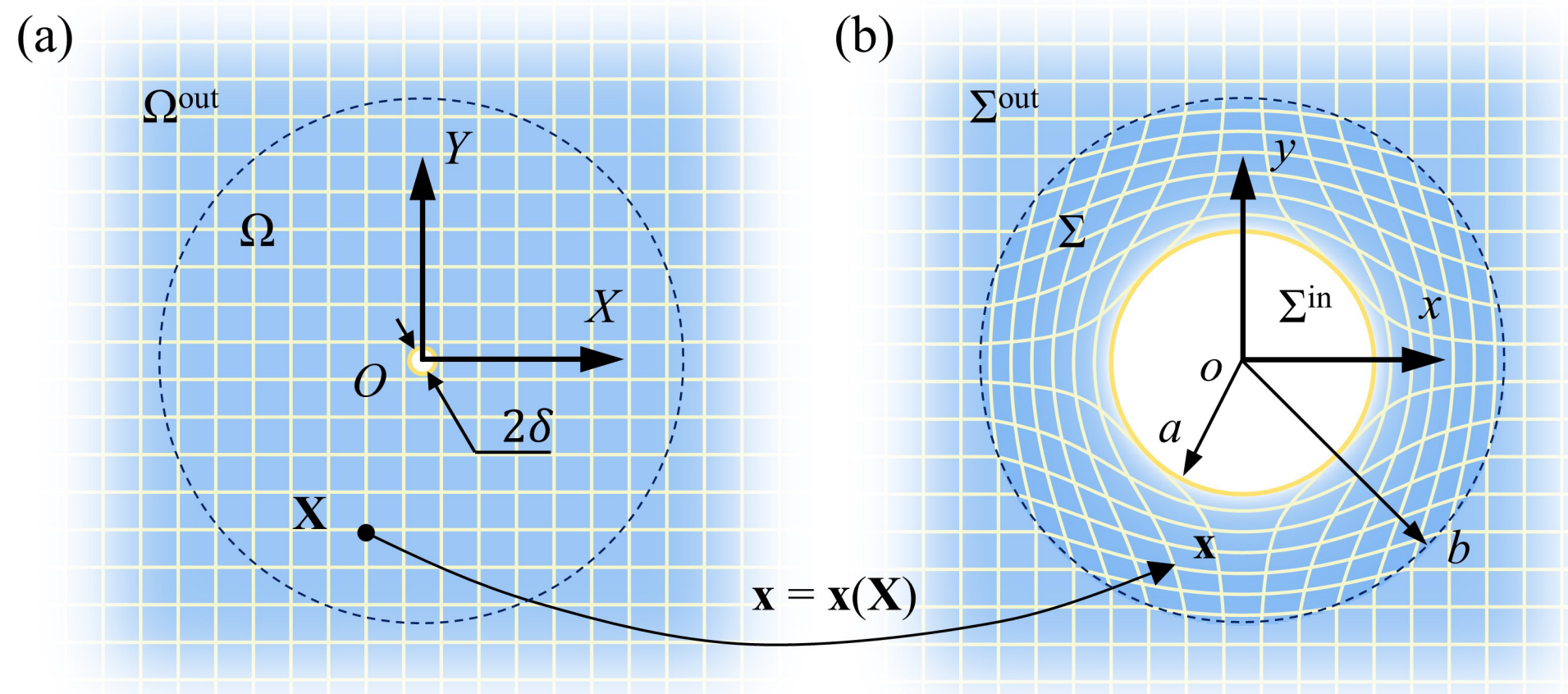

**Figure 1**. Illustration of asymmetric transformation elasticity through the design of a 2D cloak. (a) Virtual space with a background domain $\mathbf{\Omega}^{\mathrm{out}}$ and a circular domain $\mathbf{\Omega}$ with a small hole of $2\delta$ in diameter. (b) Physical space with a background domain $\Sigma^{\mathrm{out}}$, an annular cloak domain $\Sigma$, and a cloaked inner domain $\Sigma^{\mathrm{in}}$. Cartesian coordinate systems $XOY$ and $xoy$ are assigned for the virtual space and physical space, respectively. The cloak region $\Sigma$ in the physical space is mapped from the domain $\Omega$ in the virtual space through a spatial mapping $\mathbf{x} = \mathbf{x}(\mathbf{X})$.

We consider a spatial mapping $\mathbf{x} = \mathbf{x}(\mathbf{X})$ that maps a circular domain $\mathbf{\Omega}$ with a small hole in the virtual space onto a annular cloak region $\Sigma$ in the physical space. Eq.(1) then can be rewritten as the following form by a change of variable [56]

$$-\rho'\omega^2 u_i'(\mathbf{x}) = \frac{\partial \sigma_{ji}'(\mathbf{x})}{\partial x_j}, \quad \sigma_{ji}'(\mathbf{x}) = C_{ijkl}'(\mathbf{x})\frac{\partial u_i'(\mathbf{x})}{\partial x_k} \tag{2}$$

The displacement and stress in the physical space are $u_i'(\mathbf{x}) = u_i(\mathbf{X})$ and $\sigma_{ij}'(\mathbf{x}) = J^{-1}F_{ik}\sigma_{kj}(\mathbf{X})$, respectively. In the formula, the mapping gradient $F_{ij} = \partial x_i/\,\partial X_j$ and $J = \det(F_{ij})$ are defined. The mass density and elastic tensor in the cloak region $\Sigma$ are

$$\rho'(\mathbf{x}) = J^{-1}\rho_0(\mathbf{X}), \quad C_{ijkl}'(\mathbf{x}) = J^{-1}F_{im}F_{kn}C_{0mjnl}(\mathbf{X}). \tag{3}$$

The transformed elastic tensor maintains the major symmetry $C_{ijkl}' = C_{klij}'$, while violates the minor symmetry $C_{ijkl}' \neq C_{jikl}'$. As a consequence, the transformed stress tensor is not symmetric $\sigma_{ij}' \neq \sigma_{ji}'$. If one assigns materials to the cloak region $\Sigma$ according to the above parameters Eq.(3), elastic waves coming from the background domain $\Sigma^{\text{out}}$ cannot detect the cloaked inner region $\Sigma^{\text{in}}$. For clarity, we assume that the mapping gradient can be diagonalized, i.e., $\mathbf{F} = F_{11}\mathbf{e}_1 \otimes \mathbf{e}_1 + F_{22}\mathbf{e}_2 \otimes \mathbf{e}_2$, with $\mathbf{e}_1$ and $\mathbf{e}_2$ being the orthonormal bases of the principal coordinate. The material in the virtual space has an isotropic elastic tensor $C_{0ijkl} = \lambda_0\delta_{ij}\delta_{kl} + \mu_0(\delta_{ik}\delta_{jl} + \delta_{il}\delta_{jk})$. The assumption is valid for mapping for typical wave controlling function, such as cloaks. The required mass density and the elasticity matrix in Eq.(3) for the annular cloak can be simplified in the principal coordinate as

$$\rho' = \frac{\rho_0}{F_{11}F_{22}}, \quad \begin{pmatrix} \sigma'_{11} \\ \sigma'_{22} \\ \sigma'_{12} \\ \sigma'_{21} \end{pmatrix} = \begin{pmatrix} C'_{11} & C'_{12} & 0 & 0 \\ C'_{12} & C'_{22} & 0 & 0 \\ 0 & 0 & C'_{66} & C'_{69} \\ 0 & 0 & C'_{69} & C'_{99} \end{pmatrix} \begin{pmatrix} \varepsilon'_{11} \\ \varepsilon'_{22} \\ \varepsilon'_{12} \\ \varepsilon'_{21} \end{pmatrix}, \tag{4}$$

$$\begin{cases} C'_{11} = \eta(\lambda_0 + 2\mu_0), & C'_{22} = \dfrac{\lambda_0 + 2\mu_0}{\eta}, \quad C'_{12} = \lambda_0, \\ C'_{66} = \eta\mu_0, & C'_{99} = \dfrac{\mu_0}{\eta}, \qquad\quad C'_{69} = \mu_0. \end{cases} \tag{5}$$

Here, the stretching ratio $\eta = F_{11}/F_{22}$ and the asymmetric strain $\varepsilon'_{ij} = u'_{j,i}$ are defined. For a trivial stretching ratio $\eta = 1$, the required material is a conventional Cauchy material with $C'_{66} = C'_{99} = C'_{69} = \mu_0$. For this Cauchy material, an antisymmetric strain or an infinitesimal rotation ($u_{1,1} = u_{2,2} = 0$, $u_{1,2} = -u_{2,1}$) results in zero stresses and can be regarded as its easy mode. However, for a nontrivial ratio $\eta \neq 1$, the required material is an asymmetric elastic material, since the three shear modulus, $C'_{66}$, $C'_{99}$, and $C'_{69}$, are different. An infinitesimal rotation can lead to non-zero stresses in the material. It can be verified that the asymmetric material also exhibits an easy mode [22]. This easy mode is a combination of an infinitesimal rotation and a shear deformation, as indicated by $C'_{66}C'_{99} - C'^2_{69} = 0$ from Eq.(5). In the following, we will show how to achieve the asymmetric elastic properties using extremal micropolar materials with special easy modes.

*2.2 Basics of micropolar continuum theory*

We provide a brief introduction to micropolar continuum theory. In micropolar elasticity, micro-rotation DOFs, $\phi_i$, in addition to displacement DOFs, $u_i$, are introduced for each material point. The deformation is characterized by [49]

$$\varepsilon_{ij} = \frac{\partial u_j}{\partial x_i} - \epsilon_{ijk}\phi_k, \quad \kappa_{ij} = \frac{\partial \phi_j}{\partial x_i}, \tag{6}$$

where $\epsilon_{ijk}$ , $\varepsilon_{ij}$, and $\kappa_{ij}$ represents the Levi-Civita tensor, the micropolar strain tensor and the micropolar curvature tensor, respectively. All indices range from 1 to 2 for two-dimensional space and from 1 to 3 for three-dimensional space. The balance law for linear momentum and angular momentum in frequency domain read

$$-\omega^2 \rho u_i = \frac{\partial \sigma_{ji}}{\partial x_j}, \quad -\omega^2 I \phi_i = \frac{\partial m_{ji}}{\partial x_j} + \epsilon_{ijk}\sigma_{jk}, \tag{7}$$

in which, $\rho$ is the mass density, $I$ represents the micro-rotation inertia density (micro-rotation inertia per unit volume), $\sigma_{ij}$ and $m_{ij}$ stand for the stress tensor and the couple stress tensor, respectively. In Cauchy theory, the stress tensor can be proved to be symmetric from the balance law for angular momentum, owing to the absence of couple stress and micro-rotation DOFs. Here, the micropolar stress doesn't need to be symmetric and the asymmetric part is balanced by the couple stress and the micro-rotation inertia.

For infinitesimal deformations, a quadratic strain energy density function in terms of the micropolar strain and the micropolar curvature is assumed [49]

$$w = \frac{1}{2}\varepsilon_{ij} C_{ijkl} \varepsilon_{kl} + \varepsilon_{ij} D_{ijkl} \kappa_{kl} + \frac{1}{2}\kappa_{ij} A_{ijkl} \kappa_{kl}. \tag{8}$$

The constitutive law for the stress tensor and the couple stress tensor are derived by differentiating the strain energy density $w$ with respect to the strain and the curvature, respectively,

$$\sigma_{ij} = C_{ijkl}\varepsilon_{kl} + D_{ijkl}\kappa_{kl}, \quad m_{ij} = A_{ijkl}\kappa_{kl} + D_{klij}\varepsilon_{kl}. \tag{9}$$

The micropolar elastic tensor $C_{ijkl}$ and the higher-order elastic tensor $A_{ijkl}$ exhibit major symmetry, $C_{ijkl} = C_{klij}$, $A_{ijkl} = A_{klij}$, but not minor symmetry [49]. Both $C_{ijkl}$ and $A_{ijkl}$ are normal tensors, while the coupling tensor or chiral tensor $D_{ijkl}$ is a pseudo-tensor [49]. The coupling tensor is responsible for characterizing chiral effects [57, 58] and reverses its sign under space-inversion. From the thermal stability condition, $C_{ijkl}$ and $A_{ijkl}$ are positive definite and the chiral tensor $D_{ijkl}$ is bounded by the two tensors [59]. In particular, the chiral tensor must be zero, $D_{ijkl} = 0$, if the higher-order tensor vanishes, $A_{ijkl} = 0$.

### *2.3 Micropolar continuum parameters for elasticity cloak*

Now, we introduce the similarity between extremal micropolar elastic materials and asymmetric elastic materials, following a previous method [60]. We consider micropolar materials with a zero higher-order elastic tensor $A_{ijkl} = 0$. Such materials are also called reduced Cosserat media [55]. The zero higher-order tensor leads to easy deformation modes related to the micropolar curvatures. As said above, the chiral tensor also vanishes $D_{ijkl} = 0$. Both conditions can be met with a proposed micropolar metamaterial later. The constitutive equations and the dynamic equations in Eq.(7) and (9) reduce to

$$\sigma_{ij} = C_{ijkl}\frac{\partial u_l}{\partial x_k} - C_{ijkl}\epsilon_{klm}\phi_m, \quad m_{ij} = 0, \tag{10}$$

$$-\omega^2 \rho u_i = \frac{\partial \sigma_{ji}}{\partial x_j}, \quad -\omega^2 I \phi_i = \epsilon_{ijk}\sigma_{jk}. \tag{11}$$

The displacement DOFs and the micro-rotation DOFs are coupled in the constitutive law. To achieve Eq.(2) for the asymmetric elastic theory, we need to decouple the displacement and the micro-rotation. Substituting the stress tensor in Eq.(10) into the balance law of angular momentum in Eq.(11), we obtain,

$$H_{mn}\phi_n = -\epsilon_{mij}C_{ijkl}\frac{\partial u_l}{\partial x_k}, \quad H_{mn} = H_{nm} = I\omega^2\delta_{mn} - \epsilon_{mij}C_{ijkl}\epsilon_{nkl}. \tag{12}$$

We have $\det(\mathbf{H}) = 0$ for some frequencies, $\omega_r$, which correspond to the cut-off frequencies of optical modes of the micropolar material [49]. For frequencies other than the cut-off frequency $\omega \neq \omega_r$, the micro-rotation can be solved from Eq.(12). Substituting the solved micro-rotation into Eq.(10) yields

$$-\omega^2\rho u_i = \frac{\partial \sigma_{ji}}{\partial x_j}, \quad \sigma_{ij} = C_{ijkl}^{\text{eff}}\frac{\partial u_l}{\partial x_k}, \quad C_{ijkl}^{\text{eff}} = C_{ijkl} + H_{nm}^{-1}C_{ijrs}\epsilon_{rsn}\epsilon_{mpq}C_{pqkl}. \tag{13}$$

One can see that the micropolar Eq.(13) exactly matches the equation of the asymmetric elasticity Eq.(2). This means that the special micropolar material can be treated as an effective asymmetric elastic material. Alternatively, we can say that micropolar theory can provide a rigorous mathematical theory for asymmetric elastic materials. The effective asymmetric elastic tensor $C_{ijkl}^{\text{eff}}$ is frequency dependent or dispersive. We remark that Eq.(13) for balance of linear momentum is only one part of the complete dynamics equation for the micropolar material. It is this part that shares the same form with Eq. (2) for asymmetric transformation elasticity. A complete description of the dynamics of the micropolar material must be supplemented by the balance law for angular momentum Eq.(12). Otherwise, we will face the same issue as the asymmetric elastic theory, i.e., the unbalanced angular momentum caused by asymmetric stresses.

By comparing the elastic parameters, $C_{ijkl}^{\text{eff}}$, with asymmetric elastic parameters given in Eq.(4) and (5), we obtain the following micropolar parameters that can mimic the asymmetric elastic parameters

$$\begin{cases} \rho = \dfrac{\rho_0}{F_{11}F_{22}}, \quad I > 0, \quad A_{ijkl} = 0, \quad D_{ijkl} = 0, \\ C_{11} = \eta(\lambda_0 + 2\mu_0), \quad C_{22} = \dfrac{(\lambda_0 + 2\mu_0)}{\eta}, \quad C_{12} = \lambda_0, \\ C_{66} = \eta\mu, \quad C_{99} = \dfrac{\mu}{\eta}, \quad C_{69} = \mu, \quad \mu = \dfrac{\mu_0 I\omega^2\eta}{\mu_0(\eta - 1)^2 + I\omega^2\eta}. \end{cases} \tag{14}$$

The micro-rotation inertia parameter $I > 0$ can be chosen arbitrarily. The required micropolar material has an easy mode resulting from $C_{66}C_{99} - C_{69}^2 = 0$. It is noted that the three shear parameters, $C_{66}$, $C_{99}$, and $C_{69}$, depend on the desired working frequency. If the micro-inertial parameter $I$ is very large, the required parameters becomes roughly frequency independent. In such a situation, a designed micropolar material can approximate an asymmetric elastic material over a broad frequency range. In fact, if an external torque is supplied to keep zero micro-rotations of a micropolar material, the micropolar material can be regarded as a nondispersive asymmetric elastic material. This is the same as a previous 2D broadband elastic cloak, where the rotation of the inner block in each unit cell is forced to be zero [22]. For realizing three-dimensional asymmetric transformation, the required micropolar parameters are given in **Appendix A**. In the following, we demonstrate how to achieve such extremal micropolar material.

## 3. Design of the extremal micropolar material

### *3.1 Deformation modes of extremal micropolar materials*

In this section, we discuss easy deformation modes of the above extremal micropolar media. The discussion can provide a guideline for designing easy modes. The required 2D extremal micropolar material for cloaking exhibit two types of easy deformation modes. The first type is related to the vanishing higher-order elastic tensor $\mathbf{A}=\mathbf{0}$. From the constitutive law, $\mathbf{A}=\mathbf{0}$ means that the gradient of micro-rotation should not induce couple stress. In other words, a nonzero micro-rotation of a material particle should not induce micro-rotation of its neighboring particles [55]. For a metamaterial model based on mass blocks and linear Hooke's springs, as shown in next subsection, it is required that all springs must be coupled to geometry centers of those mass blocks.

As we have noted above, the second type of easy modes is represented by micropolar strain, resulting from $C_{66}C_{99}-C_{69}^2=0$ in Eq. (14). Before getting into the detail, we first provide a schematic of all possible deformation modes (not necessary be easy modes) related to the strain tensor. In general, we can decompose a micropolar strain tensor (2D case) into the following four parts, corresponding to four basic deformation modes

$$\boldsymbol{\varepsilon}=\nabla\mathbf{u}-\phi\boldsymbol{\varepsilon}^{\mathrm{a}}=\bar{\varepsilon}\boldsymbol{\varepsilon}^{\mathrm{o}}+\varepsilon_{\mathrm{d}}\boldsymbol{\varepsilon}^{\mathrm{d}}+\varepsilon_{\mathrm{s}}\boldsymbol{\varepsilon}^{\mathrm{s}}-\varphi\boldsymbol{\varepsilon}^{\mathrm{a}}, \tag{15}$$

in which,

$$\begin{cases}\boldsymbol{\varepsilon}^{\mathrm{o}}=\mathbf{e}_1\otimes\mathbf{e}_1+\mathbf{e}_2\otimes\mathbf{e}_2, & \bar{\varepsilon}=\dfrac{u_{1,1}+u_{2,2}}{2},\\ \boldsymbol{\varepsilon}^{\mathrm{d}}=\mathbf{e}_1\otimes\mathbf{e}_1-\mathbf{e}_2\otimes\mathbf{e}_2, & \varepsilon_{\mathrm{d}}=\dfrac{u_{1,1}-u_{2,2}}{2},\\ \boldsymbol{\varepsilon}^{\mathrm{s}}=\mathbf{e}_1\otimes\mathbf{e}_2+\mathbf{e}_2\otimes\mathbf{e}_1, & \varepsilon_{\mathrm{s}}=\dfrac{u_{1,2}+u_{2,1}}{2},\\ \boldsymbol{\varepsilon}^{\mathrm{a}}=\mathbf{e}_1\otimes\mathbf{e}_2-\mathbf{e}_2\otimes\mathbf{e}_1, & \varphi=\phi-\dfrac{u_{2,1}-u_{1,2}}{2},\end{cases} \tag{16}$$

with $\mathbf{e}_1$ and $\mathbf{e}_2$ representing two orthogonal bases of the coordinate system. One can verify that the four tensors, $\boldsymbol{\varepsilon}^{\mathrm{o}}$, $\boldsymbol{\varepsilon}^{\mathrm{d}}$, $\boldsymbol{\varepsilon}^{\mathrm{s}}$, and $\boldsymbol{\varepsilon}^{\mathrm{a}}$ are orthogonal to each other. $\bar{\varepsilon}\boldsymbol{\varepsilon}^{\mathrm{o}}$, $\varepsilon_{\mathrm{d}}\boldsymbol{\varepsilon}^{\mathrm{d}}$, and $\varepsilon_{\mathrm{s}}\boldsymbol{\varepsilon}^{\mathrm{s}}$ in together represents the symmetric part of the micropolar strain tensor, i.e., $\bar{\varepsilon}\boldsymbol{\varepsilon}^{\mathrm{o}}+\varepsilon_{\mathrm{d}}\boldsymbol{\varepsilon}^{\mathrm{d}}+\varepsilon_{\mathrm{s}}\boldsymbol{\varepsilon}^{\mathrm{s}}=(\nabla\mathbf{u}+\mathbf{u}\nabla)/2$. The asymmetric part of the strain tensor is exclusively represented by $\varphi\boldsymbol{\varepsilon}^{\mathrm{a}}$. We refer $\varphi$ as the pure rotation [61] because it represents the difference between the micro-rotation, $\phi$, and the displacement-related macro-rotation $(\nabla\mathbf{u}:\boldsymbol{\varepsilon}^{\mathrm{a}})/2$. We schematically represent in Fig. 2 the four deformation modes based on an infinitesimal square element. As expected, $\bar{\varepsilon}$ corresponds to a hydrostatic deformation, omnidirectional expansion or omnidirectional compression. Both $\varepsilon_\mathrm{d}$ and $\varepsilon_\mathrm{s}$ represent a pure shear mode though the deformation for $\varepsilon_\mathrm{d}$ is quite different from that of $\varepsilon_\mathrm{s}$ at first sight (compare (ii) and (iii) in Fig. 2). The mode for $\varepsilon_\mathrm{d}$ becomes the same as $\varepsilon_\mathrm{s}$ if one represent $\varepsilon_\mathrm{d}$ in a coordinate that is rotate by $45^\mathrm{o}$. The pure rotation mode, $\varphi$, is also schematically represented.

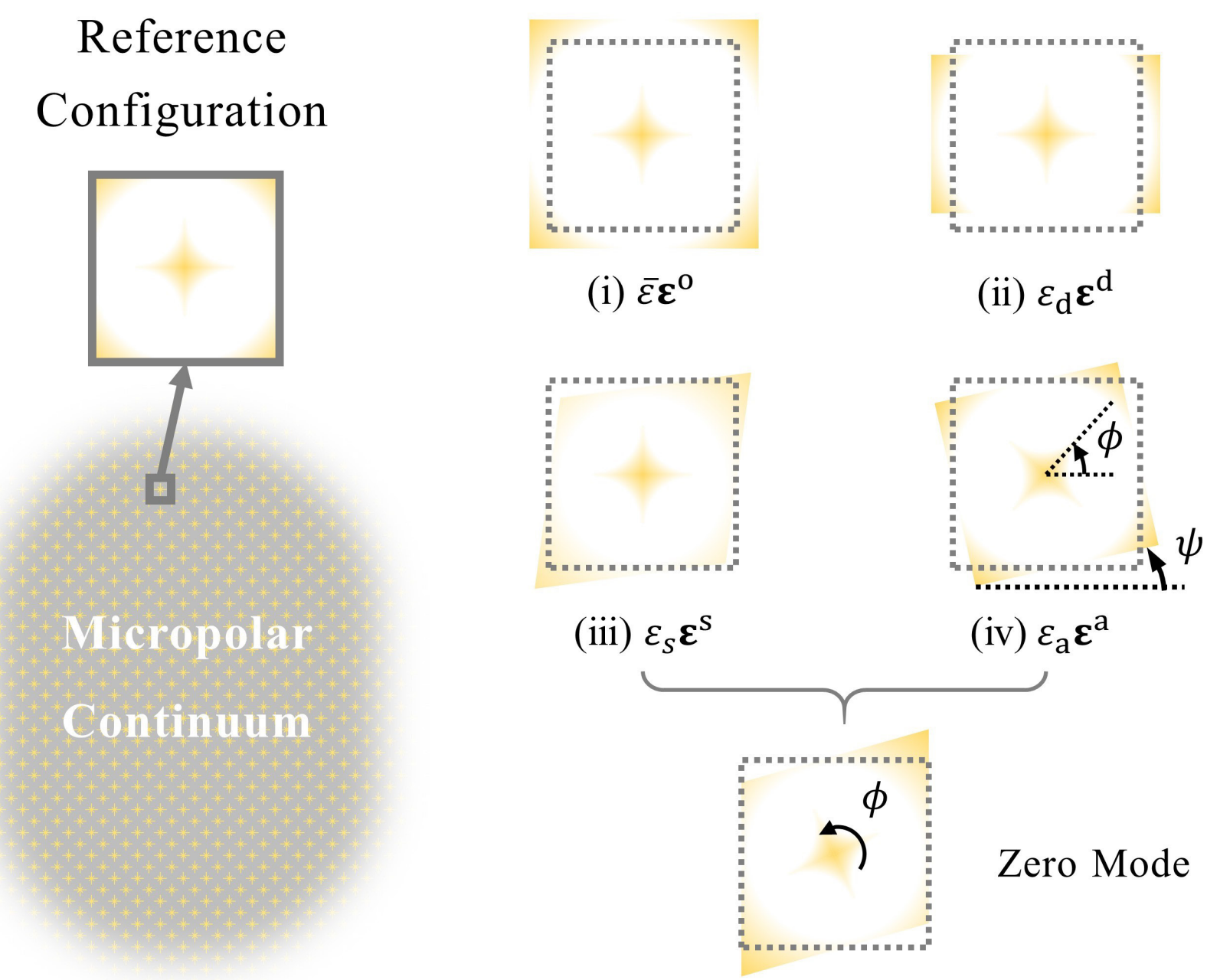

**Figure 2**. Four basic deformation modes represented by the micropolar strain tensor, $\boldsymbol{\varepsilon}$. A Cartesian coordinate with two orthonormal bases, $\mathbf{e}_1$ and $\mathbf{e}_2$, is defined. A general micropolar strain $\boldsymbol{\varepsilon}$ can be decomposed into four basic deformation modes: (i) the hydrostatic deformation $\bar{\varepsilon}\boldsymbol{\varepsilon}^\mathrm{o} = (u_{1,1} + u_{2,2})(\mathbf{e}_1 \otimes \mathbf{e}_1 - \mathbf{e}_2 \otimes \mathbf{e}_2)/2$, (ii) the first pure shear mode $\varepsilon_\mathrm{d}\boldsymbol{\varepsilon}^\mathrm{d} = (u_{1,1} - u_{2,2})(\mathbf{e}_1 \otimes \mathbf{e}_1 - \mathbf{e}_2 \otimes \mathbf{e}_2)/2$, (iii) the second pure shear mode $\varepsilon_\mathrm{s}\boldsymbol{\varepsilon}^\mathrm{s} = (u_{1,2} + u_{2,1})(\mathbf{e}_1 \otimes \mathbf{e}_2 + \mathbf{e}_2 \otimes \mathbf{e}_1)/2$, and (iv) the pure rotation mode $\varphi\boldsymbol{\varepsilon}^\mathrm{a} = (\phi - \psi)(\mathbf{e}_1 \otimes \mathbf{e}_2 - \mathbf{e}_2 \otimes \mathbf{e}_1)$, with $\psi = (u_{2,1} - u_{1,2})/2$ being the displacement-induced macro-rotation. For the extremal micropolar material in Eq.(14)(15), the easy deformation mode is a combination of a shear and the pure rotation, as shown here.

The above four modes themselves can be easy modes or their combinations can be easy modes. The easy mode of the extremal micropolar material with the parameters in Eq.(14) belongs to the latter case. The easy mode can be identified by analyzing the strain energy density

$$w = \frac{1}{2}\boldsymbol{\varepsilon}:\mathbf{C}:\boldsymbol{\varepsilon} = P\bar{\varepsilon}^2 + Q\varepsilon_\mathrm{d}^2 + R\bar{\varepsilon}\varepsilon_\mathrm{d} + \frac{\mu}{2\eta}[(1+\eta)\varepsilon_\mathrm{s} + (1-\eta)\varphi]^2,$$
$$P = \frac{(1+\eta)^2\lambda_0 + 2(1+\eta^2)\mu_0}{2\eta}, \quad Q = \frac{(1-\eta)^2\lambda_0 + 2(1+\eta^2)\mu_0}{2\eta}, \quad R = \frac{(\eta^2-1)(\lambda_0 + 2\mu_0)}{2\eta}. \tag{17}$$

From the last term in the strain energy density function, we obtain the easy mode as a linear combination of a shear and a pure rotation, i.e., $\bar{\varepsilon} = \varepsilon_\mathrm{d} = 0$, and $\varepsilon_\mathrm{s}/\varphi = (\eta - 1)/(\eta + 1)$. Therefore, the following proposed metamaterial must allow this easy mode.

*3.2 An extremal micropolar metamaterial model*

Here, we propose a discrete metamaterial model (Fig. 3(a)) to realize the above extremal micropolar material. Two lattice vectors of the metamaterial are defined as $\mathbf{a_1} = \{l, -h\}^\mathrm{T}$ and $\mathbf{a_2} = \{l, +h\}^\mathrm{T}$, respectively. Each unit cell contains a mass block (yellow rectangle) with mass $m$ and length $d$. The width of the mass block is assumed to be negligible. These mass blocks are coupled to their neighbors by linear Hooke's springs. Each mass block has two translational DOFs and one rotational DOF, which correspond to the displacement DOFs and the micro-rotation DOF, respectively. As the width of the mass blocks is negligible, one can image that the rotation of any mass block doesn't cause rotation of its neighbors. This is consistent with the requirement by a vanishing higher-order elastic tensor. The angle $\gamma$, the blue springs with spring constant $k/2$, and the vertical cyan springs with spring constant $K$ enable us to adjust the anisotropy of the effective elastic tensor. Furthermore, the gray springs with spring constant $2s$ serve to tune $C_{12}$.

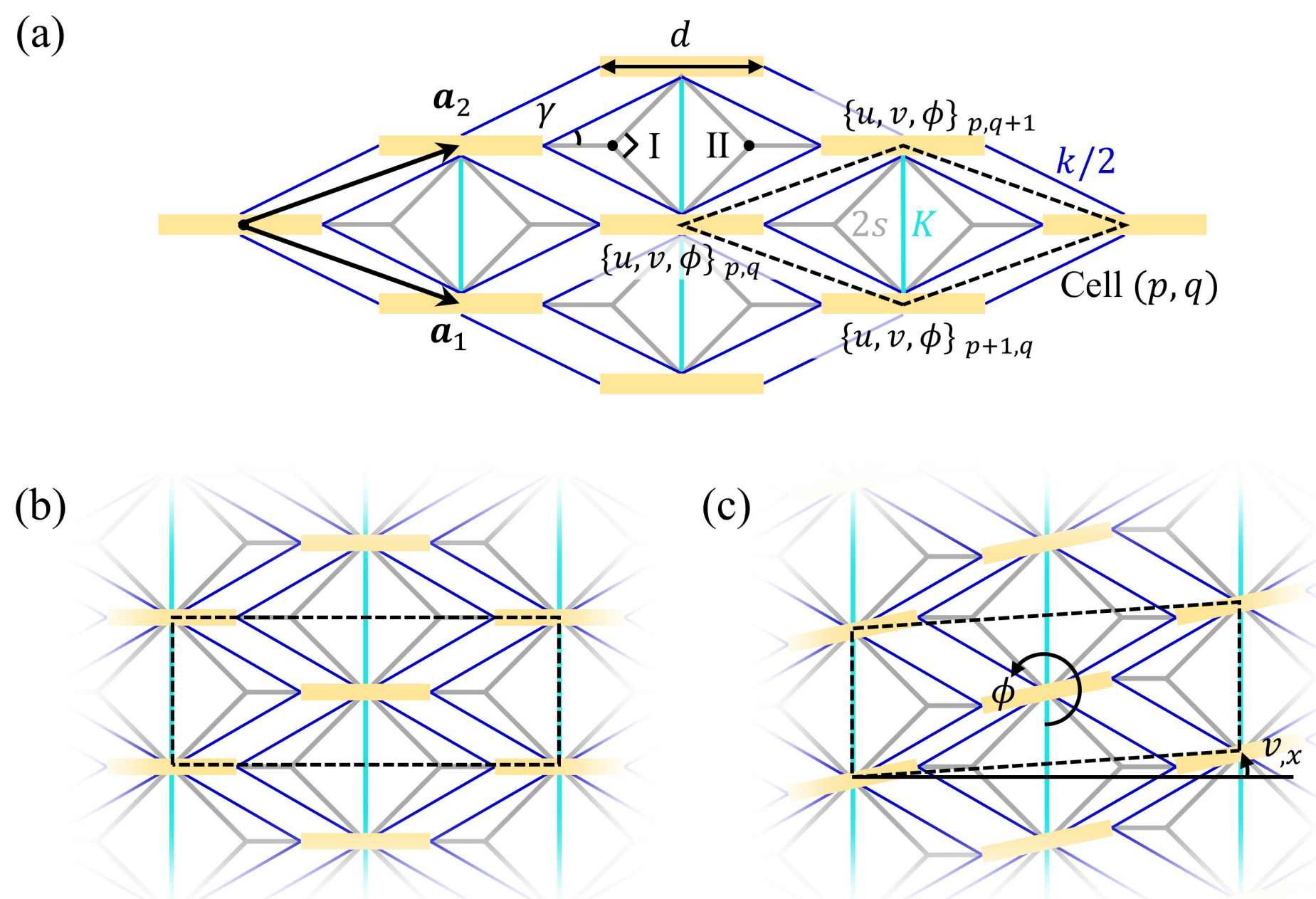

**Figure 3**. A metamaterial model for realizing extremal micropolar materials. (a) The metamaterial is consisted of rigid mass blocks (yellow rectangles) and massless springs (blue, gray and cyan lines). Geometry parameters and spring constants are indicated. The gray springs have two massless common nodes, I and II. (b) A rectangular unit cell (highlighted by the dashed box) of the metamaterial in (a). (c) Illustration of an easy mode of the metamaterial within the linear approximation. The mode is composed of shear and pure rotation.

We consider a rectangular unit cell (Fig. 3(b)) to illustrate the easy mode of the metamaterial more intuitively. First, an affine displacement represented by $u_{,x} = u_{,y} = v_{,y} = 0$, $v_{,x} > 0$ is assumed for all mass blocks. Then, all mass blocks are rotated by the same angle $\phi$. The total deformation is indicated in Fig. 3(c). It can be checked that all springs exhibit zero stretching if $\phi d/2 = l v_{,x}$ and we thus obtain an easy mode of the metamaterial. According to Eq.(16), the easy mode is represented by $\bar{\varepsilon} = \varepsilon_{\mathrm{d}} = 0$, $\varepsilon_{\mathrm{s}} = v_{,x}/2$, and $\varphi = \phi - v_{,x}/2 = v_{,x}(2l/d - 1/2)$, which is exactly a combination of shear and pure rotation, as indicated by Eq.(17). It should be noted that the metamaterial has a stretch ratio $\eta = 2l/(2l - d) > 1$. One can simply rotate the metamaterial by 90° if an asymmetric elastic material with $\eta < 1$ is needed.

In the following, we derive effective micropolar continuum parameters of the proposed metamaterial following a previous procedure [61]. We denote the displacement and rotation of each mass block as $\mathbf{u}_{p,q} = \{u_{p,q}, v_{p,q}\}^{\mathrm{T}}$ and $\phi_{p,q}$, respectively. We assume all displacements and rotations to be infinitesimal, i.e., in a linear elastic range. The kinetic energy of the unit cell $(p,q)$ can be expressed as

$$K_{p,q} = \frac{1}{2} m \left|\dot{\mathbf{u}}_{p,q}\right|^2 + \frac{1}{2} j \dot{\phi}_{p,q}^2. \tag{18}$$

The moment of inertia of the mass block with negligible thickness is $j = md^2/12$. The displacements of the left and right end-points of the mass block in the unit cell $(p,q)$ are expressed as

$$u_{p,q}^{\mathrm{L}} = \left\{u_{p,q}, v_{p,q} - \frac{\phi_{p,q} d}{2}\right\}^{\mathrm{T}}, \quad u_{p,q}^{\mathrm{R}} = \left\{u_{p,q}, v_{p,q} + \frac{\phi_{p,q} d}{2}\right\}. \tag{19}$$

The displacements of the hidden massless nodes I and II (see Fig. 3(a)) are assumed to be $\mathbf{u}_{p,q}^{\mathrm{I}}$ and $\mathbf{u}_{p,q}^{\mathrm{II}}$ respectively. We obtain the potential energy, $U_{p,q}$, for the unit cell $(p,q)$ by summing up the elastic energies of the 11 springs inside the unit cell. The expression is neglected here due to lengthy expression. Then, we derive the Lagrangian of the system

$$L = \sum_{p,q} \left(K_{p,q} - U_{p,q}\right). \tag{20}$$

The governing equations of the system are obtained as

$$\frac{\partial}{\partial t}\left(m \dot{u}_{p,q}\right) = \frac{\partial L}{\partial u_{p,q}}, \quad \frac{\partial}{\partial t}\left(m \dot{v}_{p,q}\right) = \frac{\partial L}{\partial v_{p,q}}, \quad \frac{\partial}{\partial t}\left(j \dot{\phi}_{p,q}\right) = \frac{\partial L}{\partial \phi_{p,q}}. \tag{21}$$

For the hidden DOFs $\mathbf{u}_{p,q}^{\mathrm{I}}$ and $\mathbf{u}_{p,q}^{\mathrm{II}}$, we have

$$\frac{\partial L}{\partial u_{p,q}^{\mathrm{I}}} = 0, \quad \frac{\partial L}{\partial v_{p,q}^{\mathrm{I}}} = 0, \quad \frac{\partial L}{\partial u_{p,q}^{\mathrm{II}}} = 0, \quad \frac{\partial L}{\partial v_{p,q}^{\mathrm{II}}} = 0, \tag{22}$$

Substituting the displacements of the hidden nodes solved from Eq.(22) into Eq.(21) yields the final governing equations in terms of the DOFs of the mass blocks.

The above derived governing equation is a discrete version. We consider the following Taylor series in order to obtain a continuum micropolar model. Since each mass block is only coupled to its nearest neighbors (see Fig. 3(a)), we expand $u_{p+m,q+n}$ with $m, n = -1, 0, +1$ to a second order

$$u_{p+m,q+n} = u_{p,q} + \left(\frac{\partial u}{\partial \mathbf{x}}\right)^{\mathrm{T}} \mathrm{d}\mathbf{x}_{m,n} + \frac{1}{2} \mathrm{d}\mathbf{x}_{m,n}^{\mathrm{T}} \frac{\partial^2 u}{\partial \mathbf{x}^2} \mathrm{d}\mathbf{x}_{m,n} + O\left(\left|\mathrm{d}\mathbf{x}_{m,n}\right|^3\right), \tag{23}$$

where $\mathrm{d}\mathbf{x}_{\mathrm{m,n}} = m\mathbf{a}_1 + n\mathbf{a}_2$ . $v_{p+m,q+n}$ and $\phi_{p+m,q+n}$ can be expanded similarly. Substitution of the expansion into Eq.(21) yields the following micropolar elasticity equations

$$\begin{cases} \rho \ddot{u} = C_{11} u_{,xx} + C_{99} u_{,yy} + (C_{12} + C_{69}) v_{,xy} + (C_{99} - C_{69}) \phi_{,y}, \\ \rho \ddot{v} = C_{66} v_{,xx} + C_{22} v_{,yy} + (C_{12} + C_{69}) u_{,xy} + (C_{69} - C_{66}) \phi_{,x}, \\ I \ddot{\phi} = A_{11} \phi_{,xx} + A_{22} \phi_{,yy} + 2A_{12} \phi_{,xy} + (C_{66} - C_{69}) v_{,x} + (C_{69} - C_{99}) u_{,y} + (2C_{69} - C_{66} - C_{99}) \phi, \end{cases} \tag{24}$$

where the parameters are,

$$\begin{cases} \rho = \dfrac{m}{S_{\text{cell}}}, \quad I = \dfrac{j}{S_{\text{cell}}}, \quad S_{\text{cell}} = 2lh, \quad A_{11} = A_{12} = A_{21} = A_{22} = 0, \\ C_{11} = \dfrac{l}{h}(s + k\cos^2\gamma), \quad C_{22} = \dfrac{h}{l}(s + k\sin^2\gamma + 2K), \quad C_{12} = s + k\sin\gamma\cos\gamma, \\ C_{66} = \dfrac{l}{h}k\sin^2\gamma, \quad C_{99} = \dfrac{h}{l}k\cos^2\gamma, \quad C_{69} = k\sin\gamma\cos\gamma. \end{cases} \tag{25}$$

As anticipated, the effective micropolar medium indeed has a zero higher-order elastic tensor $\mathbf{A} = \mathbf{0}$ and the condition $C_{66}C_{99} - C_{69}^2 = 0$ is also satisfied. In order to obtain the desired extremal micropolar parameters in Eq.(14), we derive the following model parameters

$$\begin{cases} m = \dfrac{\rho_0 S_{\text{cell}}}{F_{11}F_{22}}, \\ s = \lambda_0 - \mu, \quad k = \dfrac{\mu}{\sin\gamma\cos\gamma}, \quad K = \dfrac{(\lambda_0 + 2\mu_0)\cot\gamma - k\sin^2\gamma - (\lambda_0 - \mu)}{2}, \\ \gamma = \operatorname{arccot}\left(\dfrac{\sqrt{(\lambda_0 - \mu)^2 + 4\mu(\lambda_0 + 2\mu_0)} - (\lambda_0 - \mu)}{2\mu}\right). \end{cases} \tag{26}$$

It should be mentioned that the geometry requirement of $\arctan(h/l) < \gamma < \pi/4$ (see Fig. 3(a)) imposes limitation to the available extremal micropolar parameters. Specifically, the Poisson's ratio of the background material must be positive $\nu > 0$, otherwise, the required micropolar parameters in Eq.(14) cannot be achieved with the proposed metamaterial. For a previously proposed metamaterial [26] to achieve asymmetric transformation, the Poisson's ratio of the background material is limited to be $\nu < 0.25$.

*3.3 Verification of the extremal micropolar metamaterial*

Next, we demonstrate that the proposed micropolar metamaterial can be used for controlling elastic waves based on asymmetry transformation elasticity. For simplicity, we consider a stretch transformation denoted by $F_{11} = 2/3$ and $F_{22} = 1$. Elastic parameters of the back ground Cauchy material are taken as $\lambda_0 = 2\mu_0 = 110$ GPa and $\rho_0 = 2.4 \times 10^3$ kg/m$^3$, arbitrarily. We consider an operating frequency $f_{\text{ext}} = \omega_{\text{ext}}/(2\pi) = 6300$ Hz. The metamaterial cell should be sufficiently small compared to the shear wavelength in the background material, i.e., $l \ll c_0/f_{\text{ext}}$. Here, we choose the unit cell size $l = (c_0/f_{\text{ext}})/2N$ with $N = 10$. The geometry and material parameters for the metamaterial model follows Eq.(26),

$$\begin{cases} m = 3.52 \text{ kg}, \\ k = 4.34 \times 10^9 \text{ N/m}, \quad s = 1.08 \times 10^{11} \text{ N/m}, \quad K = 1.62 \times 10^{11} \text{ N/m}, \\ l = 3.80 \text{ cm}, \quad h = 1.29 \text{ cm}, \quad d = 2.53 \text{ cm}. \end{cases} \tag{27}$$

We obtain dispersion relations for the micropolar continuum and the metamaterial by substituting the harmonic wave assumption $\mathbf{u} = \hat{\mathbf{u}}\exp(\mathrm{i}(\mathbf{k}\cdot\mathbf{x} - \omega t))$ and the Bloch-Floquet ansatz $\mathbf{u}_{p+m,q+n} = \mathbf{u}_{p,q}\exp(\mathrm{i}\mathbf{k}\cdot(m\mathbf{a}_1 + n\mathbf{a}_2))$, $\mathbf{u}_{p,q} = \hat{\mathbf{u}}_{p,q}\exp(-\mathrm{i}\omega t)$ into Eq.(24) and Eq.(21), respectively [61]. There are two acoustic bands related to the two displacement DOFs. In addition, we obtain a micro-rotation dominated optical branch, with a cut-off frequency, $\omega_{\mathrm{r}} = 3.89 \times 10^4\ \mathrm{rad/s}$. The three bands for the continuum model (red lines in Fig. 4(a)) and the microstructure lattice (chain lines in Fig. 4(a)) are in good agreement in the long-wavelength limit, i.e., around the Γ point. Since the Taylor expansion Eq.(23) is more accurate for small wave numbers, we observe a larger discrepancy between the two results for larger wavenumbers, especially along two principal directions of the micropolar material, i.e., the ΓX direction and ΓY direction. Along the two principal directions, the shear branch for the continuum model exhibits zero frequency due to the above explained easy mode of the metamaterial. The longitudinal mode and the optical mode are completely decoupled (see Fig. 4(b)) and the two branches are degenerate at a certain wavenumber along the two directions. However, for wave propagating along directions other than the two principal directions, such as the ΓK direction and ΓM direction, the transverse mode, longitudinal mode and optical mode all couple together, leading to a non-zero-frequency transverse branch and a lifted degeneracy between the longitudinal branch and the optical branch. As result, the degeneracy along the two principal directions forms a Dirac cone [62]. It is interesting that the micropolar continuum calculation perfectly reproduce the Dirac cone of the metamaterial. The micropolar elasticity can potentially provide an continuum route to investigate Dirac physics [63].

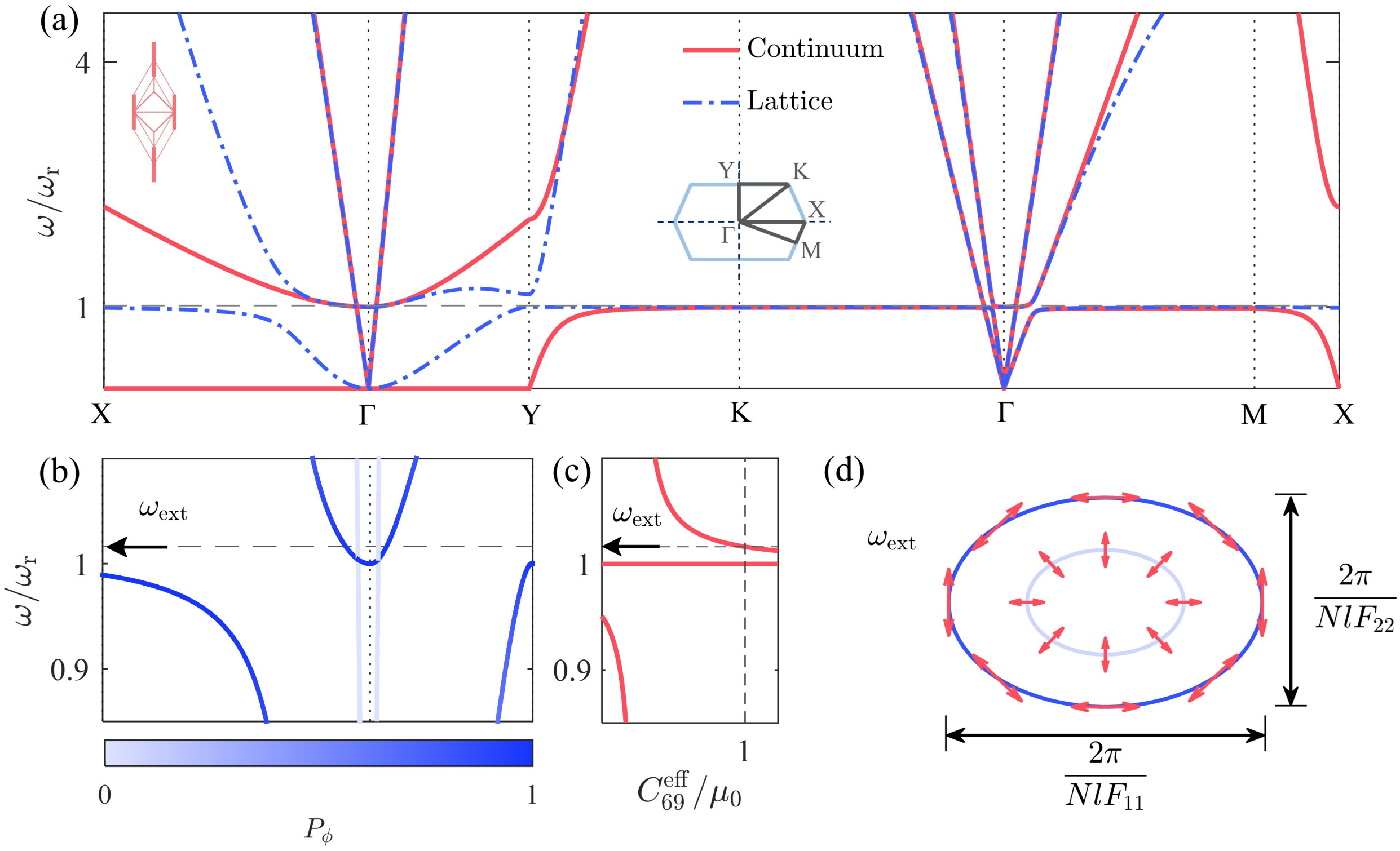

**Figure 4**. (a) Dispersion bands calculated from the micropolar continuum model (red solid line) and the proposed microstructure lattice (blue chain line). The insets show a unit cell of the metamaterial (top-left) and the first Brillouin zone (center). (b) Zoomed-in view of the band structure of the lattice near the operating frequency $\omega_{\mathrm{ext}}$ along the path X − Γ − Y. The contribution of micro-rotation to each mode, defined as $P_\phi = j\hat{\phi}^2/(m\hat{u}^2 + m\hat{v}^2 + j\hat{\phi}^2)$, are encoded by color. Dark blue indicates strong participation of micro-rotation while light blue means micro-rotation is negligible in the mode. (c) Effective relative shear modulus $C_{69}^{\mathrm{eff}}$. (d) Iso-frequency contour of the metamaterial at the operating frequency $\omega_{\mathrm{ext}}$. Displacement polarization of the modes are represented by red double arrows.

The dashed gray line in Fig. 4(b) indicates the operating frequency $\omega_{\text{ext}}$ of the designed metamaterial. The effective shear modulus $C_{69}^{\text{eff}}$ at the operating frequency $\omega_{\text{ext}}$ matches the shear modulus of the background material (see Fig. 4(c)). At the operating frequency, $\omega_{\text{ext}}$, the iso-frequency contour of the metamaterial are two ellipses with the same aspect ratio as the stretch ratio $\eta = 2/3$. This is necessary for perfectly matching the wave property of the background material, as demonstrated in a previous asymmetric elastic metamaterial [26]. A notable difference between our current metamaterial and the previous one [26] is that we here exploit an acoustical mode and an optical mode, while the previous design relies on two optical modes.

Next, we verify in COMSOL Multiphysics that the wave property of the above extremal micropolar metamaterial. Specifically, we consider a metamaterial slab, with a thickness of $48h$, sandwiched in the background Cauchy medium (Fig. 5(a)). The background Cauchy medium is modeled by using the Solid Mechanics module. The mass blocks and the springs in the metamaterial are modeled by using the Beam module in COMSOL Multiphysics, respectively. We couple the background Cauchy continuum and the metamaterial slab at their interface by imposing pointwise displacement continuity to corresponding nodes of the two regions. The simulation frequency is set to be the operating frequency $f_{\text{ext}} = 6300\text{Hz}$. We show in Fig. 5(a) the simulated displacement fields for a Gaussian beam incident onto the micropolar matching layer at an angle of $45°$. The incident wave is perfectly transmitted through the matching layer, with negligible wave scattering or reflection. The wave fields inside the matching layer are compressed by a factor of $F_{11} = 2/3$ as expected [26]. The energy flow direction and the wavevector inside the matching layer are in agreement with the iso-frequency curve analysis (see Fig. 5(b) and 5(c)). The matching layer should be valid for arbitrary incidence angles. This can be verified by simulating an incident cylindrical wave, which consists of plane waves along all propagation directions in 2D space. Nearly perfect transmission for the incident cylindrical wave is observed, either for the transverse case (Fig. 5(d)) or the longitudinal case (Fig. 5(f)).

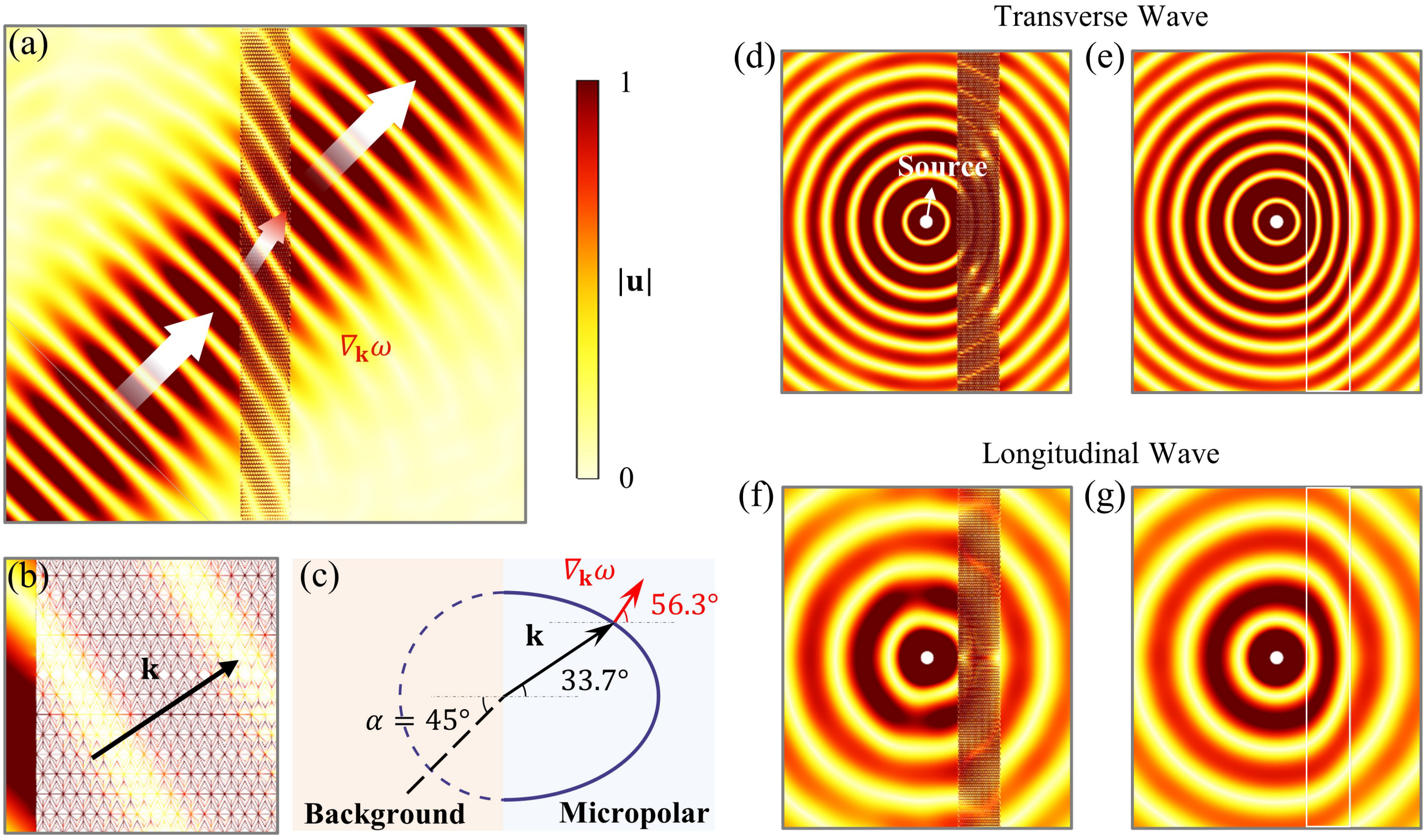

**Figure 5.** Simulation of a micropolar matching layer with different incident waves. (a) Displacement fields for a transverse polarized Gaussian beam obliquely incident onto the micropolar matching layer. Colors stand for the displacement magnitude $|\mathbf{u}|$. Arrows are for energy flow directions. (b) Zoomed-in view and the wave vector. (c) Iso-frequency contour of the background Cauchy material and the micropolar metamaterial at the operating frequency, $\omega_{\text{ext}}$. (d), (e) Results for the metamaterial and an effective micropolar continuum with an incident transverse cylindrical wave. The region enclosed by the white rectangle box represents the micropolar continuum medium. (f) (g) Same as (d), (e) but for an incident longitudinal cylindrical wave.

We also simulate the micropolar matching layer by using the derived effective micropolar continuum parameters in Eq.(25). The micropolar elasticity equation is implemented in weak form in COMSOL Multiphysics [59]. In the simulation, we assume that both the surface traction and the surface couple stress are continuous at the interface separating the background Cauchy medium and the micropolar continuum. The continuity equations can be derived by analyzing the balance of linear momentum and angular momentum for an infinitely thin strip containing the interface. The displacement is also supposed to be continuous at the interface, resulting from a perfect bonding between the two media, while the micro-rotation of the micropolar medium at the interface is not constrained. The micropolar continuum calculations (Fig. 5(e)(g)) show a perfect transmission of the incident wave. The above simulations validate the proposed extremal micropolar metamaterial.

## 4. Design and verification of 2D micropolar elasticity cloak

Now, we start the design and verification of an elastic cloak based on the proposed extremal micropolar metamaterial. We choose a linear mapping (see Fig. 1(a)), $g(r) = b(r-a)/(b-a) + \delta(r-b)(a-b)$ in a polar coordinate for the cloak design. The function maps a circular region with a hole, with radius $\delta$, in the virtual space onto an annular cloak with inner radius $a$ and outer radius $b$ in the physical space. The mapping gradient becomes $\mathbf{F} = 1/g'(r)\mathbf{e}_{\mathrm{r}} \otimes \mathbf{e}_{\mathrm{r}} + r/g(r)\mathbf{e}_{\theta} \otimes \mathbf{e}_{\theta}$, with $\mathbf{e}_{\mathrm{r}}$ and $\mathbf{e}_{\theta}$ being the unit bases along the $r$- and $\theta$-direction of the polar coordinate. We choose the same background Cauchy material as before, $\lambda_0 = 2\mu_0 = 110$ GPa and $\rho_0 = 2.4 \times 10^3$ kg/m$^3$. The geometry parameters are chosen as $b = 2a = 1$ m and $\delta = a/5$. The required micropolar continuum parameters are obtained from Eq. (14) and the microstructure parameters are further derived from Eq. (26), which depend on the radial location $r$.

The resulting cloak is consisted of in total $N_r = 123$ microstructures layers along the radial direction and $N_\theta = 84$ sectors along the $\theta$ direction. We illustrate in Fig. 6(a) the construction process of the metamaterial cloak. The metamaterial unit cells in the cloak are arranged such that the long axis of the mass blocks is along the $\theta$-direction of the polar coordinate since the required $\eta < 1$. The metamaterial unit cell in the outermost layer (or the first layer) has the largest size. We choose $l_1 = (c_0/f_{\mathrm{ext}})/2N$ with $N = 10.16$ for the first layer to ensure the long wavelength condition. For all unit cells in the cloak (see Fig. 6(a)), we have replaced the rectangular mass blocks by polyline-shaped blocks to maintain conformation with the polar coordinate. All springs or beams are coupled to mass blocks at corresponding edges. Geometric parameters of the unit cells at the location, $r_n$, are as indicated. The linear mapping results in a constant length of the mass block, $d = 2\pi b(a-\delta)/N_\theta(b-\delta)$. We have used two angles, $\gamma_n^{\mathrm{o}}$ and $\gamma_n^{\mathrm{i}}$, (see Fig. 6(a)) for each layer. The outer one is chosen as $\gamma_n^{\mathrm{o}} = \arctan(h_{n-1}/(\eta(r_n)l_n))$, while the inner one $\gamma_n^{\mathrm{i}}$ is derived from the required micropolar properties at that location. The metamaterial cloak is constructed layer by layer from the first layer with the iteration relation $r_n = r_{n-1} - h_{n-1}$ and $h_n = \eta(r_n) l_n \tan(\gamma_n^{\mathrm{i}})$.

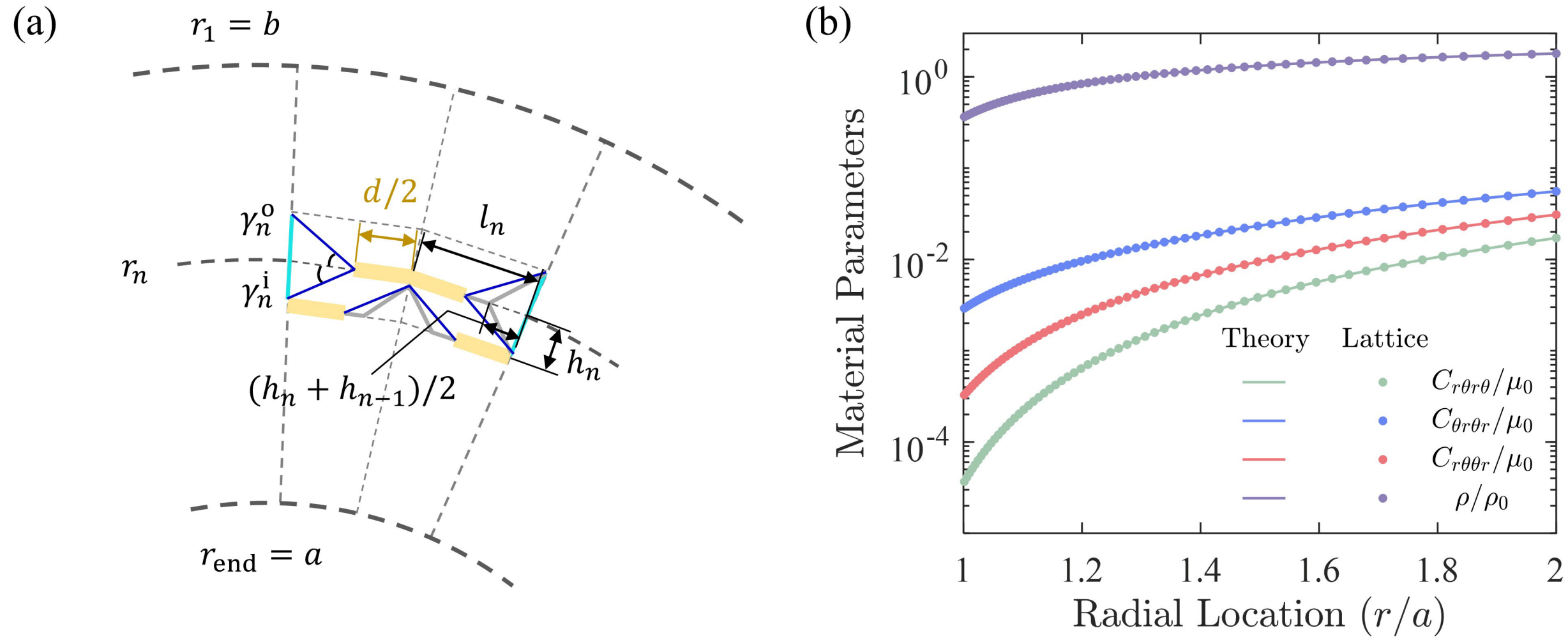

**Figure 6.** Microstructure elastic cloak based on extremal micropolar metamaterials. (a) Sketch of the construction process of the microstructure cloak. Geometric parameters of the unit cells in the *n*th layer are marked. We number the outermost layer as layer 1. (b) Effective micropolar parameters (dots) of each layer along the radial direction for the designed microstructure cloak. We only show the three shear modulus and mass the density for clarity. The designed material properties are in perfect agreement with the micropolar parameters (solid lines) required by cloaking theory.

We show in Fig. 6(b) the distribution of the effective micropolar properties of the designed cloak along the radial direction. Due to a very large number of metamaterial layers along the radial direction, the effective parameters (dots) of the metamaterial cloak approximates the continuous parameters (solid lines) required by transformation theory very well. This leads to a good approximation of the microstructure cloak to the theoretical design, as demonstrated below. As we have shown previously, the designed micropolar metamaterial unit cell has three easy modes. A finite structure made of the same cell will have deformation modes that cost zero strain energy. The designed metamaterial cloak is instead mechanically stable due to the gradient design of the metamaterial cells.

We model the microstructure cloak in COMSOL Multiphysics like the simulations in previous Section 3.3. The background material and the metamaterial are modeled by using the Solid Mechanics Module and the Beam Module, respectively. We apply a traction free boundary condition to the inner surface of the cloak. We verify the cloaking performance of the designed elastic cloak by considering a plane wave incident from the left side. The entire simulation domain is surrounded by the Perfectly Matched Layers in COMSOL Multiphysics to avoid undesired reflections from boundaries. We first show the simulated displacement fields for a transverse plane wave (Fig. 7(a)) and a longitudinal plane wave (Fig. 7(e)) onto a circular void. The void induces rather pronounced scattering to the left side and the right side of the void. In sharp contrast, the incident plane wave smoothly propagates through the cloak to the forward region when the void is covered by the microstructure cloak (Figs. 7(b)(f)). The cloak demonstrates a very satisfying cloaking performance. The displacement fields for the microstructure cloak exhibit quite good consistency with the simulation based on effective micropolar medium (Fig. 6(c)(g)). We remark that a small radius $\delta = a/5$ is adopted for the cloak design and weak scattering of the cloak is inevitable. As can be seen, the cloaking performance of the cloak is rather close to the theoretical target of a small hole with radius $\delta$ (Fig. 6(d)(h)).

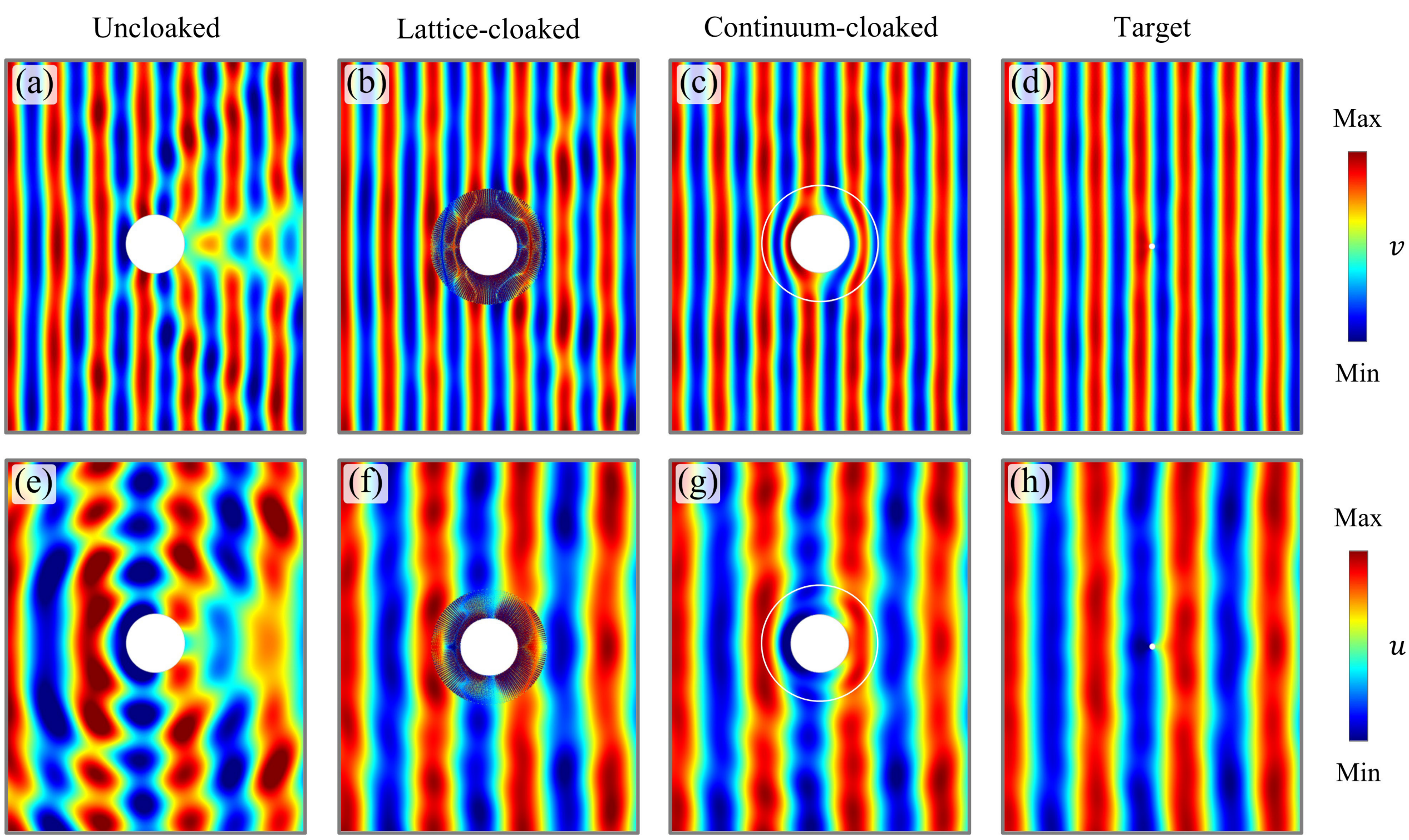

**Figure 7**. Verification of the designed cloak based on extremal micropolar metamaterials. (a) Simulated displacement fields, $v$ (vertical component of the displacements), for a transverse plane wave incident from the left side onto a void (inner circular white region) with radius $a$. (b) Same as (a) but with the void covered by the designed microstructure cloak, with inner radius $a$ and outer radius $b = 2a$. (c) Same as (b) but with micropolar effective-medium for the cloak (enclosed by white circle). (d) Same as (a) but for a small void with radius $\delta = a/5$, which is adopted for the design of the cloak. (e) – (h) Same as (a) – (d) but for displacement fields, $u$ (horizontal component of the displacements), with an incident longitudinal plane wave.

We calculate the total scattering cross-section (TSCS) for the cloak [9] under incident plane transverse and longitudinal waves. The displacement and stress fields are decomposed into the incident and scattered components, $\mathbf{u} = \mathbf{u}^{\text{in}} + \mathbf{u}^{\text{sc}}$ and $\boldsymbol{\sigma} = \boldsymbol{\sigma}^{\text{in}} + \boldsymbol{\sigma}^{\text{sc}}$ , respectively. The incident displacement fields for transverse and longitudinal waves are respectively, $\mathbf{u}^{\text{in}} = \hat{\mathbf{u}}_0 \exp(\mathrm{i}\omega x/c_{\text{T}})$ and $\mathbf{u}^{\text{in}} = \hat{\mathbf{u}}_0 \exp(\mathrm{i}\omega x/c_{\text{L}})$, where $c_{\text{T}}$ and $c_{\text{L}}$ are corresponding phase velocities in the background media.

By choosing an annular integral path $\partial\Gamma$ enclosing the cloak region, the total scattered power of the cloak can be obtained by integrating over this boundary:

$$E_{\text{sc}} = \oint \frac{1}{2} \text{Re}\left[\sigma_{ij}^{\text{sc}} \dot{\bar{u}}_j^{\text{sc}}\right] n_i ds = -\frac{\omega}{2} \text{Im} \oint n_i \sigma_{ij}^{\text{sc}} \bar{u}_j^{\text{sc}} ds \,, \tag{28}$$

where $\mathbf{n}$ is the outward normal vector to $\partial\Gamma$, and the overbar denotes the complex conjugate.

The incident power density for a plane wave is given as:

$$P_{\text{in}} = \frac{1}{2} \text{Re}\left[\sigma_{ij}^{\text{in}} \dot{\bar{u}}_j^{\text{in}}\right] n_i = -\frac{\omega}{2} \text{Im}\left[n_i \sigma_{ij}^{\text{in}} \bar{u}_j^{\text{in}}\right], \tag{29}$$

with $\mathbf{n} = (1 \quad 0)^{\text{T}}$. From these, the dimensionless TSCS is computed as $E_{\text{sc}}/(2bP_{\text{in}})$, where $b$ is the radius of the cloak region.

The scattered displacement fields under the excitation of transverse and longitudinal incident waves are shown in Figure 8. The lattice-cloak and the continuum-cloak significantly reduces the scattering outside the cloak compared to the unloaded cases in Fig. 8(b)(e). Quantitatively, the TSCS of the microstructure cloak and the continuum-cloak reduces are reduced to one-tenth of the TSCS of a bare rigid scatter, approaching the TSCS of a target void with radius $a/5$. These results clearly validate the efficiency of the designed elastic cloak based on extremal micropolar metamaterials.

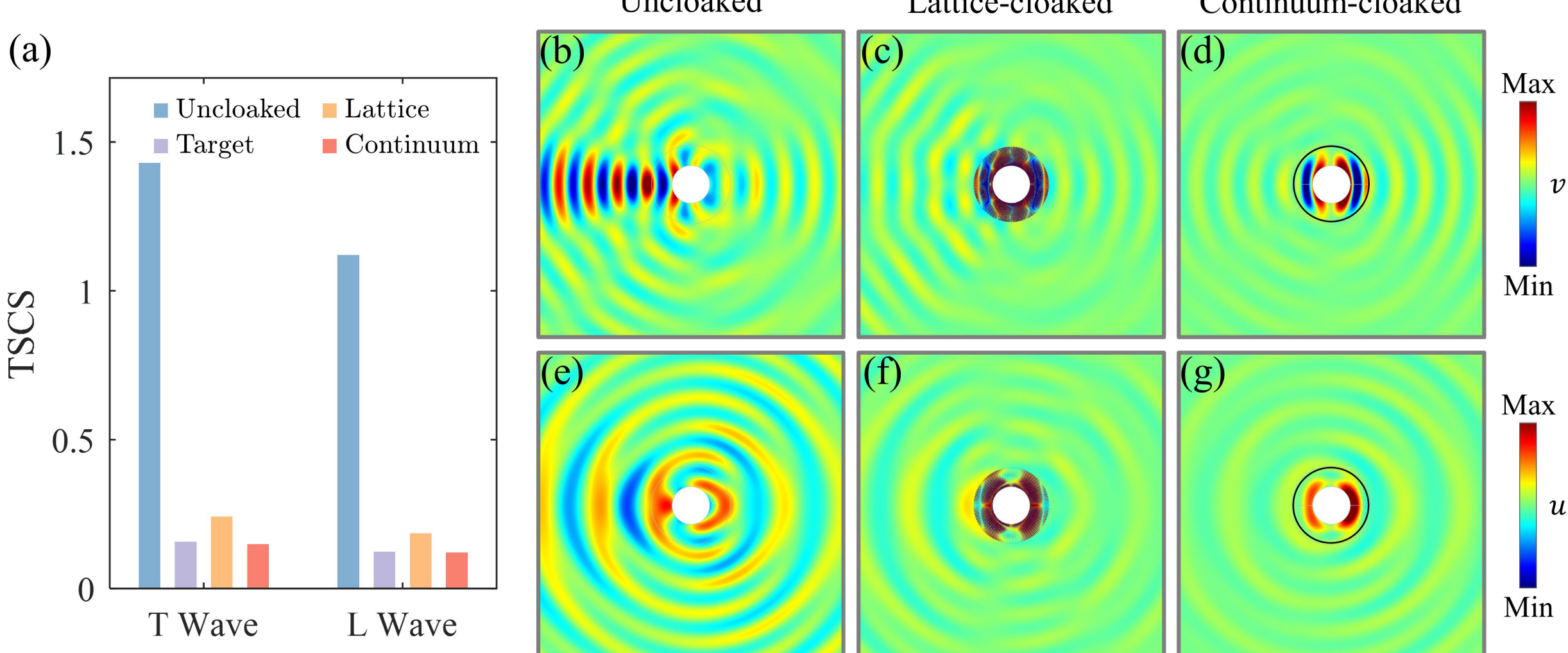

**Figure 8**. The TSCS and scattering displacement fields for both the lattice-cloak and the continuum-cloak. (a) The TSCS for each case is calculated from the scattered displacement fields shown in (b) - (g). The target TSCS is corresponding to the case illustrated in Fig. 7(d)(h). (b) Simulated scattered displacement fields, $v^{\mathrm{sc}}$, for a transverse plane wave incident onto a void (inner circular white region) with radius $a$. (c) Same as (b) but with the void covered by the designed microstructure cloak. (d) Same as (c) but with an effective-medium micropolar cloak. (e) – (g) Same as (b) – (d) but for scattered displacement fields, $u^{\mathrm{sc}}$, with an incident longitudinal plane wave.

## 5. Conclusion

Asymmetric elastic theory and asymmetric elastic materials were conceived to achieve elastic cloaks by using transformation methods nearly two decades ago. The theory shares the same math equations as Cauchy elasticity except that the elastic tensor only has major symmetry. As a result, the theory can lead to asymmetric stresses, such as by an infinitesimal rotation. However, the asymmetric elastic theory does not take into account the unbalanced angular momentum caused by the asymmetric stresses. Therefore, the transformation theory based on asymmetric elastic materials still remains controversial today, though elastic cloaks guided by the theory have already been experimentally demonstrated.

Our paper has focused on micropolar elastic theory, which allows asymmetric stresses and the asymmetric part is balanced by the couple stress and micro-rotation inertia. Specially, we are interested in extremal micropolar materials with zero higher-order elastic tensor, also termed as reduced Cosserat media. In such media, the micro-rotation degrees of freedom can be eliminated from the governing equation for linear momentum. The resulting dynamic equation for displacements becomes the same as the asymmetric elastic theory, but with an effective dispersive asymmetric elastic tensor. Therefore, extremal micropolar elastic materials are capable of realizing elastic cloaks based on transformation methods. To verify our finding, we have proposed a metamaterial to realize the extremal micropolar parameters required for cloaking. An elastic cloak has been further designed based on the metamaterial and numerically simulated. The excellent cloaking performance has justified our above finding.

Our finding also indicates that micropolar elastic theory can provide a rigorous theory for asymmetric elastic materials. This may facilitate the study of other interesting properties in asymmetric elastic materials. Extremal Cauchy materials, e.g., pentamode materials, have shown great potentials in wave control in past few years. Our work also shed light on unusual properties in extremal micropolar materials and may stimulate further exploration.

## Appendix A

The mapping gradient is assumed to be diagonal in the principal coordinate, i.e., $\mathbf{F} = F_{11}\mathbf{e}_1 \otimes \mathbf{e}_1 + F_{22}\mathbf{e}_2 \otimes \mathbf{e}_2 + F_{33}\mathbf{e}_3 \otimes \mathbf{e}_3$, with $\mathbf{e}_1$, $\mathbf{e}_2$, and $\mathbf{e}_3$ being the orthonormal bases of the principal coordinate. The required micropolar elastic parameters in the principal coordinate are

$$\rho = \frac{\rho_0}{F_{11}F_{22}F_{33}}, \quad I > 0, \quad A_{ijkl} = 0, \quad D_{ijkl} = 0, \tag{A1}$$

$$\begin{cases} C_{11} = (\lambda_0 + 2\mu_0)\dfrac{F_{11}}{F_{22}F_{33}}, \quad C_{22} = (\lambda_0 + 2\mu_0)\dfrac{F_{22}}{F_{11}F_{33}}, \quad C_{33} = (\lambda_0 + 2\mu_0)\dfrac{F_{33}}{F_{11}F_{22}}, \\ C_{12} = \dfrac{\lambda_0}{F_{33}}, \quad C_{13} = \dfrac{\lambda_0}{F_{22}}, \quad C_{23} = \dfrac{\lambda_0}{F_{11}}, \\ C_{44} = \dfrac{\mu_0 I\omega^2 F_{22}^2}{\mu_0(F_{22} - F_{33})^2 + I\omega^2 F_{11}F_{22}F_{33}}, \quad C_{77} = \dfrac{\mu_0 I\omega^2 F_{33}^2}{\mu_0(F_{22} - F_{33})^2 + I\omega^2 F_{11}F_{22}F_{33}}, \quad C_{47} = \sqrt{C_{44}C_{77}}, \\ C_{55} = \dfrac{\mu_0 I\omega^2 F_{11}^2}{\mu_0(F_{11} - F_{33})^2 + I\omega^2 F_{11}F_{22}F_{33}}, \quad C_{88} = \dfrac{\mu_0 I\omega^2 F_{33}^2}{\mu_0(F_{11} - F_{33})^2 + I\omega^2 F_{11}F_{22}F_{33}}, \quad C_{58} = \sqrt{C_{55}C_{88}}, \\ C_{66} = \dfrac{\mu_0 I\omega^2 F_{11}^2}{\mu_0(F_{11} - F_{22})^2 + I\omega^2 F_{11}F_{22}F_{33}}, \quad C_{99} = \dfrac{\mu_0 I\omega^2 F_{22}^2}{\mu_0(F_{11} - F_{22})^2 + I\omega^2 F_{11}F_{22}F_{33}}, \quad C_{69} = \sqrt{C_{66}C_{99}}. \end{cases} \tag{A2}$$

## Data and code availability

The data that support the plots within this paper are publically available on the open access data repository of Zenodo [https://dx.doi.org/10.5281/zenodo.13377851].

## Acknowledgements

We thank Hongkuang Zhang for valuable discussions. This work is support by the National Natural Science Foundation of China (Grant Nos. 11632003, 11872017, 11972083, 11972080) and partially by the Alexander von Humboldt foundation.

## Author contributions

D.S. designed the metamaterial. D.S. and Y.C. performed the theory analysis and numerical simulations. All authors contributed to the writing, review and revision of the paper. Y.C. and G.K. supervised the project.

## Conflict of interests

Authors declare that they have no competing interests.